\documentclass[11pt,letterpaper]{article}
\pdfoutput=1
\usepackage{jheppub}
\usepackage{bbm}
\usepackage{mathrsfs}
\usepackage{slashed}
\usepackage{caption}
\usepackage{epstopdf}
\usepackage[normalem]{ulem}
\usepackage[bottom]{footmisc}
\usepackage{subcaption}
\usepackage{bbold}
\usepackage{titlesec}
\usepackage{threeparttable}
\usepackage{booktabs}
\usepackage{changepage}
\usepackage[utf8]{inputenc}
\usepackage{dsfont}

\usepackage{grffile}
 
\usepackage{graphicx}  
\usepackage{dcolumn}   
\usepackage{bm}        
\usepackage{amssymb}   
\usepackage{setspace}
\usepackage{amsmath, amssymb, setspace}
\usepackage{array}
\usepackage{booktabs}
\usepackage{caption}
\usepackage{indentfirst}
\usepackage{float}
\usepackage{lmodern}
\usepackage{multirow}
\usepackage{soul}
\usepackage[normalem]{ulem}

\usepackage{braket}
\usepackage{comment}

\usepackage[draft]{pgf}

\usepackage{adjustbox} 
\newcommand{\Eq}[1]{Eq.~(\ref{#1})}

\newlength{\myimageoversize}
\newsavebox{\myimage}

\usepackage[most]{tcolorbox}

\usepackage{empheq}

\tcbset{colback=white!10!white, colframe=black!50!black, 
        highlight math style= {enhanced, 
            colframe=black,colback=white!10!white,boxsep=0pt}
        }

\titleformat{\subsection}
 {\normalfont\fontsize{12}{17}\itshape}{\thesubsubsection}{1em}{}
 
\title{\huge{Cosmological Dependence of Sterile Neutrino Dark Matter With Self-Interacting Neutrinos}}

\author[a]{Carlos Chichiri,}
\author[b]{Graciela B. Gelmini,}
\author[b,c]{Philip Lu}
\author[d]{and Volodymyr Takhistov}

\affiliation[a]{Department of Physics and Astronomy, California State University, Northridge\\
 Northridge, CA 91330, USA}

\affiliation[b]{Department of Physics and Astronomy, University of California, Los Angeles\\
 Los Angeles, CA 90095-1547, USA}

\affiliation[c]{Center for Theoretical Physics, Department of Physics and Astronomy, \\
Seoul National University, Seoul 08826, Korea}

\affiliation[d]{Kavli Institute for the Physics and Mathematics of the Universe (WPI),\\University of Tokyo, Chiba 277-8583, Japan}
\emailAdd{carlos.chichiri.87@my.csun.edu}
\emailAdd{gelmini@physics.ucla.edu}
\emailAdd{philiplu11@gmail.com}
\emailAdd{volodymyr.takhistov@ipmu.jp}

\abstract{
Unexplored interactions of neutrinos could be the key to understanding the nature of the dark matter (DM). In particular, active neutrinos with new self-interactions can produce keV-mass sterile neutrinos that account for the whole of the DM through the Dodelson-Widrow mechanism for a large range of active-sterile mixing values. This production typically occurs before Big-Bang Nucleosynthesis (BBN) in a yet uncharted era of the Universe. 
~We assess how the mixing range for keV-mass sterile neutrino DM is affected by the uncertainty in the early Universe pre-BBN cosmology.
This is particularly relevant for identifying the viable parameter space of sterile neutrino searches allowed by all astrophysical limits, as well as for cosmology, since the detection of a sterile neutrino could constitute the first observation of a particle providing information about the pre-BBN epoch. We find that the combined uncertainties in the early Universe cosmology and neutrino interactions significantly expand the allowed parameter space for sterile neutrinos that can constitute the whole of the DM.
}

\begin{document}
\preprint{IPMU21-0076}
 \maketitle
\flushbottom

\section{Introduction}

Sterile (right-handed) neutrino, a Standard Model (SM) gauge singlet of mass $m_s$ that can couple to the SM active neutrinos $\nu_{\alpha}$, where $\alpha = e, \mu, \tau$, constitutes a viable $\mathcal{O}$(keV) mass-scale dark matter (DM) candidate\footnote{Other sterile neutrino parameter space ranges can also be of interest, e.g. heavier sterile neutrinos in connection with Hubble parameter tension~\cite{Fuller:2011qy,Gelmini:2019deq,Gelmini:2020ekg}.} (see e.g. Ref.~\cite{Boyarsky:2018tvu}). 
Sterile neutrinos are an essential ingredient of the seesaw mechanism~\cite{Yanagida:1979as,GellMann:1980vs,Yanagida:1980xy,Minkowski:1977sc}, one of the simplest methods for generating observed~\cite{Fukuda:1998mi} small masses for active neutrinos.  For simplicity, we consider a $\nu_s$ that mixes with one active neutrino through a mixing angle $\sin \theta$, which we assume to be $\nu_\mu$.

Without additional interactions, sterile neutrinos are produced in the early Universe through active-sterile flavor oscillations and collisional processes.
Sterile neutrinos produced via non-resonant oscillations, through the Dodelson-Windrow (DW) mechanism~\cite{Dodelson:1993je}, are already strongly constrained by astrophysical observations, especially X-ray data, as the dominant DM component (e.g.~\cite{Palazzo:2007gz,Horiuchi:2013noa,Perez:2016tcq,Ng:2019gch,Dessert:2018qih}).
The putative 3.5 keV X-ray emission line signal observed from galaxy clusters has also been suggested to originate from decaying sterile neutrinos of $m_s = 7.1$ keV and a mixing  $\sin^2 2\theta = 5\times10^{-11}$ assuming these sterile neutrinos constitute all of the DM abundance~\cite{Bulbul:2014sua,Boyarsky:2014jta}, although this has been challenged (e.g.~\cite{Dessert:2018qih}).

In alternative scenarios, existing constraints on sterile neutrino DM abundance can be alleviated. This can occur in the presence of a significant initial lepton asymmetry\footnote{While a lepton asymmetry is much less restricted than the baryon asymmetry, observations show that it does not need to be significant for sterile neutrino resonant production~(e.g.~\cite{Mangano:2011ip,Oldengott:2017tzj}).}, leading to resonant active-sterile oscillations~\cite{Shi:1998km}. Other possibilities, which can appear in models that are often more complicated, include sterile neutrino production in the context of inflaton decays~\cite{Shaposhnikov:2006xi}, freeze-in decays of other particles~\cite{Roland:2014vba,Asaka:2006ek}, and extended gauge symmetries~\cite{Dror:2020jzy,Bezrukov:2009th}. 

DW sterile neutrino production can be affected by additional unknown active neutrino ``non-standard interactions''(NSI), also known as ``secret interactions"~(see e.g.~\cite{Proceedings:2019qno} for review). NSI play a role in a variety of well motivated scenarios~(e.g.~\cite{Chikashige:1980qk,Gelmini:1980re,Ng:2014pca,Ioka:2014kca,Blinov:2019gcj,Bustamante:2020mep}) 
and considerable experimental efforts have been devoted to explore them. The parameter space for keV-mass sterile neutrino DM can be significantly expanded in the presence of scalar~\cite{DeGouvea:2019wpf,Kelly:2020aks} or vector~\cite{Kelly:2020pcy} mediated neutrino NSI. Such possibilities can readily appear in models associated with Majorons~\cite{Chikashige:1980qk,Gelmini:1980re} (for scalar mediated NSI)  or gauged $U(1)$ symmetries (for vector mediated NSI). This highlights that unknown neutrino interactions can modify the early Universe reaction rates and play an important role in determining the theoretical uncertainties associated with identifying viable parameter space regimes for sterile neutrino searches. Self-interactions among sterile neutrinos can also play a role in their production~\cite{Johns:2019cwc,Hansen:2017rxr}.

It is well known that cosmological history can significantly impact expectations for sterile neutrino production, their abundance and momentum distribution (e.g.~\cite{Gelmini:2004ah,Rehagen:2014vna,Gelmini:2019esj,Gelmini:2019wfp,Gelmini:2019clw,Gelmini:2020duq}). Light nuclei produced during the Big Bang Nucleosynthesis (BBN) are the earliest cosmological remnants that have been detected thus far and the cosmological history of the Universe prior to BBN (i.e. temperatures $T>$ 5 MeV) remains unknown~\cite{deSalas:2015glj, Hasegawa:2019jsa, DeBernardis:2008zz, Hannestad:2004px, Kawasaki:2000en, Kawasaki:1999na}. In many well motivated models, such as those based on moduli decay~\cite{Gelmini:2006pw,Gelmini:2006pq,Moroi:1999zb,Chen:2018uzu,Kitano:2008tk,Kawasaki:1995cy}, quintessence~\cite{Salati:2002md,Profumo:2003hq,Pallis:2005hm} or extra dimensions~\cite{Randall:1999vf,Durrer:2005dj}, the cosmological history could be different than typically assumed. It has been demonstrated~\cite{Gelmini:2019esj,Gelmini:2019wfp,Gelmini:2019clw} that sterile neutrinos constitute an excellent probe of the yet untested cosmological epoch before BBN. Conversely, this also implies that there is significant uncertainty in the viable sterile neutrino parameter space relevant for searches associated with the unknown cosmological history.

In this work we analyze the cosmological dependence of sterile neutrino DM in the presence of neutrino NSI. This simultaneous consideration of two major sources of theoretical uncertainties allows to broadly asses robustness of predictions and determine expectations for experimental sterile neutrino searches. The results of this study also outline the scope of how much sterile neutrinos can act as probes of the early Universe. While for concreteness we focus on scalar mediated neutrino NSI in the context of different cosmological models, including scalar-tensor and kination, our analysis can be readily generalized to other possibilities.
We generally consider sterile neutrinos to compose the entirety of the DM density and call this scenario ``sterile neutrino DM", although our results can be easily extrapolated to subdominant DM densities.
 
This paper is organized as follows. In Section~\ref{sec:modcos} we describe a
parametrization of the Hubble expansion rate of the Universe $H$ and define the particular pre-BBN cosmological models that we consider. In Section~\ref{sec:prod}, we outline the Boltzmann formalism for sterile neutrino production and introduce the effects of NSI. In Section~\ref{sec:limits}, we explain our NSI parameter scan and the relevant astrophysical limits for keV-mass scale sterile neutrinos as well as other searches. A technical interpretation of the results is included. In Section~\ref{sec:summary} we summarize and discuss implications of the study.
 
\section{Early Universe cosmology}
\label{sec:modcos}

A common choice for the early pre-BBN cosmological history, which we denote as ``standard" (Std),  assumes that the Universe was radiation-dominated up to temperatures  of the radiation bath   much larger than the temperature at which BBN happen. In this case, the expansion rate of the Universe $H(T)$ has at $T \gg$ MeV the same form we know at lower temperatures  (see e.g.~\cite{Kolb:1990vq})
\begin{equation} 
\label{eq:hStd}
H_{\rm Std}(T) =  \dfrac{T^2}{M_{\rm Pl}}  \sqrt{\dfrac{8 \pi^3 g_{\ast}(T)}{90}}~,
\end{equation}
where $M_{\rm Pl}= 1.22 \times 10^{19}$ GeV is the Planck mass and $g_\ast(T)$ is the energy density number of degrees of freedom  (see e.g.~\cite{Husdal:2016haj,Borsanyi:2016ksw,Drees:2015exa}). With only the SM degrees of freedom present, $g_\ast(T)$ is $g_\ast \simeq 80$ above the QCD phase transition, at $T \simeq 1$ GeV,  decreases steeply close to the QCD phase transition to $g_{\ast} \simeq 30$ at $T \simeq 200$ MeV, is $g_{\ast} = 10.75$ between $T\simeq 20$ MeV and $T =1$ MeV (when electrons and positrons become non-relativistic and annihilate), and is $g_{\ast} = 3.36$ for $T <1$ MeV.
The temperature range where we expect most of the sterile neutrino production to occur is below a few 100 MeV. Thus, as a good approximation that simplifies our calculations, we fix $g_\ast(T)$ to $g_\ast=10.75$ throughout.

In the different cosmologies we consider, the unknown cosmological evolution of the early Universe is allowed to be drastically distinct as long as it transitions to the standard cosmology at $T < 5$ MeV. Here, $T = 5$ MeV is the lower limit on the highest temperature of the standard radiation dominated era
imposed by the known late history of the Universe~\cite{deSalas:2015glj, Hasegawa:2019jsa, DeBernardis:2008zz, Hannestad:2004px, Kawasaki:2000en, Kawasaki:1999na}.
We explore cosmologies where entropy in matter and radiation is conserved and the expansion rate in the non-radiation dominated phase can be described by a phenomenological parameterization valid for $T> T_{\rm tr}$ for a large class of models~\cite{Catena:2009tm}, 
\begin{equation} \label{eq:hnStd}
    H = \eta~ \left(\dfrac{T}{T_{\rm tr}}\right)^{\beta} H_{\rm Std}~,
\end{equation}
described in terms of two real parameters, $\eta> 0$ and $\beta$, and  the transition temperature $T_{\rm tr}$, defined as the temperature below which the cosmology becomes standard.
Following our previous study of non-resonant sterile neutrino production~\cite{Gelmini:2019esj, Gelmini:2019wfp}, we consider several vastly different specific examples of cosmologies of this type discussed in the literature:
\begin{itemize}
\item the Kination model (K)~\cite{Spokoiny:1993kt,Joyce:1996cp,Salati:2002md,Profumo:2003hq,Pallis:2005hm}, with $\eta = 1$, $\beta = 1$;
\item the Scalar-Tensor model  of Ref.~\cite{Catena:2004ba} (ST1), with $\eta = 7.4 \times 10^5$, $\beta = -0.8$;
\item the Scalar-Tensor model  of Ref.~\cite{Catena:2007ix} (ST2),  with $\eta = 0.03$, $\beta = 0$.  
\end{itemize}

\begin{figure}[tb]
\begin{center}
\includegraphics[width=.5\textwidth]{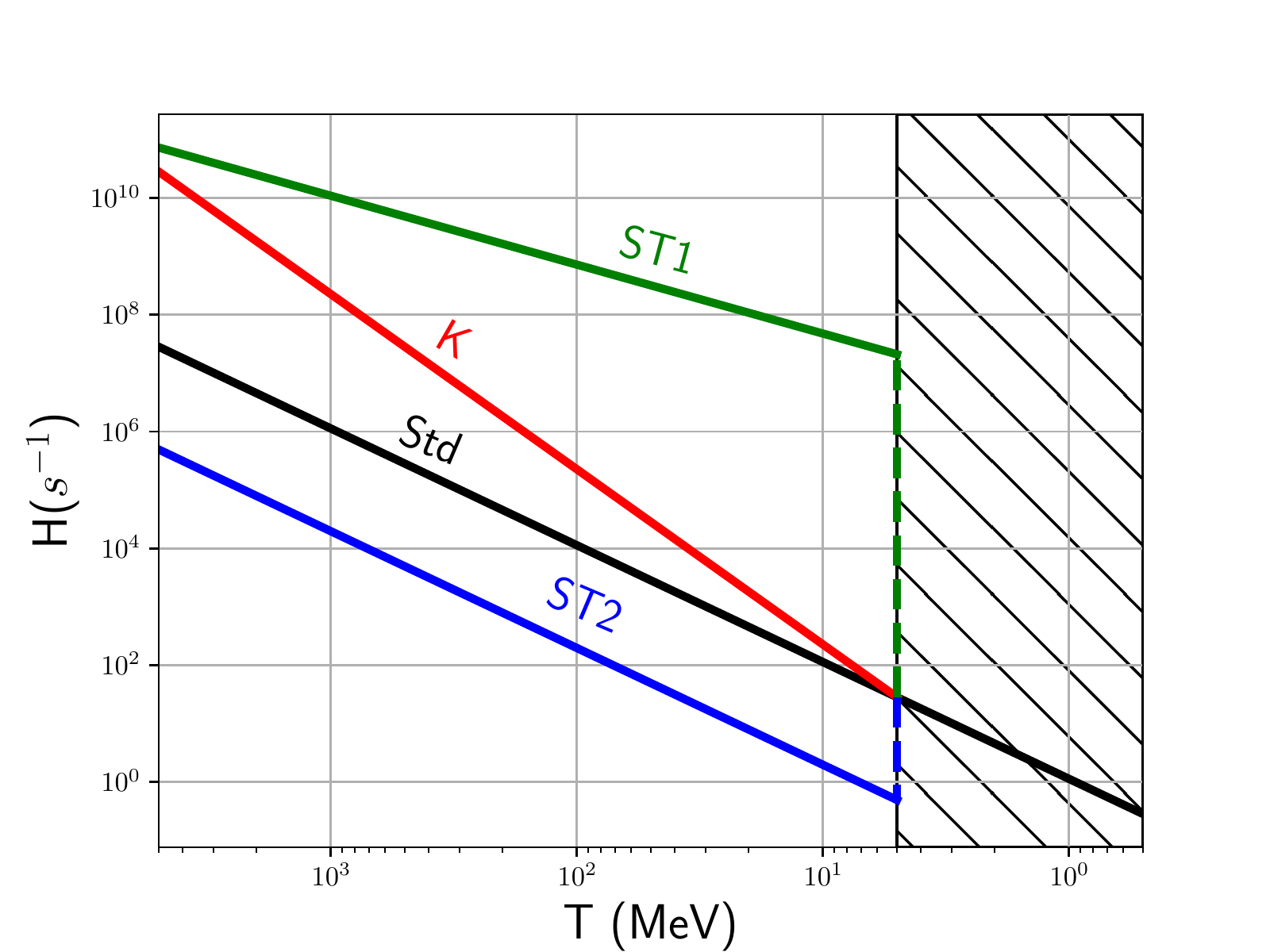}
\caption{The Hubble expansion rate $H$ as a function of the temperature $T$ of the Universe for the cosmologies considered: Std (black), K (red), ST1 (green), and ST2 (blue). All cosmologies are taken to transition to the standard cosmology at $T_{tr}=5$ MeV. \label{fig:hgraph}}
\end{center}
\end{figure}

We choose $T_{\rm tr} = 5$~MeV throughout to analyze the maximum possible extent of the cosmological effects, while ensuring that
there are no additional constraints from astrophysical observations (e.g. binary star mergers~\cite{Sakstein:2017xjx}).
For simplicity, we assume an abrupt transition of the Scalar-Tensor models into the standard cosmology at $T_{\rm tr}$ and comment later about the consequences of this choice. $H(T)$ for the four cosmologies we consider is shown in Fig.~\ref{fig:hgraph}.

\section{Sterile neutrino production with self-interacting active neutrinos}
\label{sec:prod}

The time evolution of the momentum
density function  $f_{s}(p,t)$ of sterile neutrinos  is given in terms of the equilibrium  density function of $\nu_{\alpha}$ active neutrinos $f_{\nu_{\alpha}}(p,t) = (e^{\epsilon}+1)^{-1}$ by the Boltzmann equation
~\cite{Kolb:1990vq,Abazajian:2001nj, Rehagen:2014vna}
\begin{equation}
\label{eq:boltzmann2}
    -HT\left(\frac{\partial f_{s}(E,T)}{\partial T}\right)_{E/T=\epsilon} \simeq~ \Gamma(E,T)f_{\nu_\alpha}(E,T)~,
\end{equation}
which is valid only if
 $f_{s} \ll 1$ and $f_{s} \ll f_{\nu_{\alpha}}$. Here $p$ is the magnitude of the neutrino momentum vector, the derivative in the left-hand side of the equation is computed at constant $\epsilon  = p/T$,
and $\Gamma (p,t)$ is the active-to-sterile conversion rate.
All neutrino species are relativistic when sterile neutrinos are produced, thus the energy is $E\simeq p$.

The total rate $\Gamma$ of active-to-sterile  neutrino conversion is given by the probability $\langle P_m \rangle$ of an active-sterile flavor oscillation in matter (for  mixing angle in matter $\theta_m$) times the active neutrino interaction rate $\Gamma_{\rm int}$
\begin{equation} 
\label{eq:interaction}
    \Gamma ~=~ \dfrac{1}{2}  \langle P_m (\nu_\alpha \rightarrow \nu_s) \rangle ~ \Gamma_{\rm int}   ~\simeq~ \frac{1}{4}\sin^2(2\theta_m)~ \Gamma_{\rm int}~.
\end{equation}
Following Refs.~\cite{DeGouvea:2019wpf} and~\cite{Kelly:2020pcy} (see also references therein) we consider that besides the standard weak interactions, the active neutrinos also have NSI though the exchange of a boson mediator field. This could be a complex scalar $\phi$ of mass $m_\phi$ and coupling constant $\lambda_\phi$~\cite{DeGouvea:2019wpf}, or a vector $V$ of mass $m_V$ and coupling constant $\lambda_V$~\cite{Kelly:2020pcy}. 
Hence,
\begin{equation} 
\label{eq:gamma-int}
    \Gamma_{\rm int}~=~ \Gamma_{\rm weak} + \Gamma_{\rm NSI}~.
\end{equation}
Here, $\Gamma_{\rm weak}$ is the weak interaction rate of active neutrinos with the surrounding plasma ~\cite{Abazajian:2001nj}
\begin{equation} 
\label{eq:interaction-active}
    \Gamma_{\rm weak} = d_{\alpha} G_F^2 \epsilon T^5~,
\end{equation}
 where $d_{\alpha} = 1.27$ for $\nu_e$ and $d_\alpha = 0.92$ for $\nu_{\mu}$, $\nu_{\tau}$. We are going to assume that the sterile neutrino couples mostly with $\nu_{\mu}$.

With a scalar mediator, the NSI rate of active neutrinos for $T \ll m_\phi$ is
\begin{equation}
\label{eq:interaction-off}
\Gamma_{\rm NSI}=\Gamma_\phi^{\rm low-T} (m_\phi, \lambda_\phi) =\frac{7\pi \lambda_\phi^4 \epsilon T^5}{432~ m_\phi^4}~,
\end{equation}
 while for 
 $T\gtrsim m_\phi$  it is
\begin{equation}
\label{eq:interaction-on}
\Gamma_{\rm NSI}= \Gamma_\phi^{\rm high-T} (m_\phi, \lambda_\phi) =\Gamma_\phi^{\rm on-shell}(m_\phi, \lambda_\phi) \simeq \frac{\lambda_\phi^2 m_\phi^2 }{4\pi \epsilon^2 T} \left( \ln \left( 1 + e^{w} \right) - w\rule{0mm}{4mm}\right),
\end{equation}
where $w= m_\phi^2/(4 \epsilon  T^2)$. Here \Eq{eq:interaction-off}  and \Eq{eq:interaction-on} are respectively Eq.~(5) and Eq.~(6) of Ref.~\cite{DeGouvea:2019wpf}, multiplied by a factor of 2 to take into account the charge conjugate interactions.

There are many different possibilities for the coupling of a vector mediator to active neutrinos, as explored in Ref.~\cite{Kelly:2020pcy}. To show how our results change with different vector mediated interaction models we chose two examples. First, we consider an additional neutrinophilic $U(1)_{np}$ gauge symmetry with a vector boson $Vnp$ that couples solely to active neutrinos, and, following Ref.~\cite{Kelly:2020pcy} we consider the representative case of coupling to $\nu_\mu$ only.
In this case, as shown in Sec.~III of~Ref.~\cite{Kelly:2020pcy}, the NSI interaction rate for temperatures $T \ll m_V$ is,
\begin{equation}
\label{eq:interaction-npV-low}
\Gamma_{\rm NSI}=\Gamma_{Vnp}^{\rm low-T}+ \Gamma_{Vnp}^{\rm on-shell}= 8~\Gamma_\phi^{\rm low-T}(m_V, \lambda_V) + \Gamma_{Vnp}^{\rm on-shell}~,
\end{equation}
 where  
\begin{equation}
\label{eq:interaction-npV-on-shell}
\Gamma_{Vnp}^{\rm on-shell}= \Gamma_\phi^{\rm on-shell}(m_V, \lambda_V)
\end{equation}
(with $w= m_V^2/(4 \epsilon  T^2)$) is the dominant term in Eq.~\eqref{eq:interaction-npV-low} only for $T\simeq m_V$, and for 
 $T\gtrsim m_V$, instead, 
\begin{equation}
\label{eq:interaction-npV-high}
\Gamma_{\rm NSI}=\Gamma_{Vnp}^{\rm high-T}+\Gamma_{Vnp}^{\rm on-shell}= \frac{3 \zeta(3) |\lambda_V|^4 T^3}{4\pi^3 m_V^4} + \Gamma_{Vnp}^{\rm on-shell}.
\end{equation}

Second, we consider NSI based on an additional $U(1)_{B-L}$ gauge symmetry (e.g.~\cite{Mohapatra:1980qe}) with a gauge vector boson $V_{B-L}$ that couples universally to all quark and lepton flavors.
Following Sec.~V of~Ref.~\cite{Kelly:2020pcy}), for a gauged $U(1)_{B-L}$ model, the NSI rate due to the exchange of a vector boson for  $T \lesssim m_V$ is
\begin{equation}
\label{eq:interaction-B-L-low}
\Gamma_{\rm NSI}=\Gamma_{V_{B-L}}^{\rm on-shell}+\Gamma_{V_{B-L}}^{\rm low-T} \simeq  \Gamma_{V_{B-L}}^{\rm on-shell} + 1.6~\Gamma_\phi^{\rm low-T}(m_V, \lambda_V) \times
\left\{ 
\begin{array}{cl}
22,&\hspace{1cm} 2m_\mu \leq T \\
21,&\hspace{1cm} m_\mu \leq T< 2m_\mu \\
17,& \hspace{1cm} 2m_e\leq T<m_\mu \\
16,& \hspace{1cm} m_e\leq T<2 m_e \\
12,& \hspace{1cm} T<m_e\, . \\
\end{array}\right.
\end{equation}
Here, $\lambda_V$ is the $B-L$ gauge coupling constant, and 
\begin{equation}
\label{eq:interaction-B-L-on-shell}
\Gamma_{V_{B-L}}^{\rm on-shell} \simeq \frac{\lambda_V^4 m_V^2}{96\pi^2 \gamma_V \epsilon^2 T} \left( \ln \left(1 + e^{\omega} \right) - \omega \rule{0mm}{4mm}\right) \,\times
\left\{ 
\begin{array}{cl} 
7,&\hspace{1cm} 2m_\mu \leq T \\
5,& \hspace{1cm} 2m_e \leq T<2m_\mu \\
3,& \hspace{1cm} T<2m_e \\
\end{array}\right.
\end{equation}
where $w\equiv {m_V^2}/{(4\epsilon T^2)}$, and $\gamma_V$ is the decay width of the $B-L$ gauge boson~\cite{Heeck:2014zfa}.

Fig. 1 of Ref.~\cite{Heeck:2014zfa} shows that the dominant contribution to $\gamma_V$ is due to the leptonic  decays into three active majorana neutrinos, $e^\pm$, and $\mu^\pm$. Taking into account these contributions, and neglecting  a narrow hadronic resonance (which does not significantly affect our results), we approximate the decay rate to be  $\gamma_V\simeq 7\lambda_V^2 m_V/24\pi$ (see Eq. (3) of Ref.~\cite{Heeck:2014zfa}).   
 
 For $T \gtrsim m_V$, we use $\Gamma_{\rm NSI} = \Gamma_{V_{B-L}}^{\rm on-shell}$ for the $B-L$ model.  
The NSI interaction rate for $T \gg m_V$  in this model is not needed since laboratory experiments reject light vector bosons (see e.g.  Fig.~8 of~Ref.~\cite{Kelly:2020pcy}), so that sterile neutrino DM is dominantly produced at temperatures where either $\Gamma_{V_{B-L}}^{\rm low-T}$ or $\Gamma_{V_{B-L}}^{\rm on-shell}$ is dominant, or if the maximum of production happens at larger temperatures, the NSI interactions are negligible and the production happens because of the standard weak interactions.

The mixing angle  in matter $\theta_m$ in the presence of an active neutrino  NSI is given by
\begin{equation}
\label{eq:mattermixing}
    \sin^2(2\theta_m) = \frac{\sin^2(2\theta)}{\sin^2(2\theta) +
    \dfrac{\Gamma_{\rm int}^2 \epsilon T}{2m_s^2} +
    \left[\cos(2\theta)-\dfrac{2\epsilon T}{m_s^2} (V_D+V_T+V_{\rm NSI})\right]^2}~.
\end{equation} 
In the denominator of Eq.~\eqref{eq:mattermixing}, the density potential  $V_D$
depends on the particle-antiparticle asymmetries in the background, which we assume to be negligible, thus   $V_D=0$. $V_T$ is the standard thermal potential
\begin{align}
 V_T=-B\epsilon T^5~, 
 \end{align}
where $B$ is a constant prefactor which depends on the active neutrino flavor (assumed here to be $\nu_\mu$)  
and temperature range  
and we take to be $B= 10.88\times10^{-9}$GeV$^{-4}$.

The contribution of the NSI to the thermal potential  for a scalar mediator is~\cite{Notzold:1987ik, Quimbay:1995jn, DeGouvea:2019wpf}
\begin{equation}
\label{eq:vphi}
\begin{split}
V_{\rm NSI}= V_\phi(E, T, m_\phi, \lambda_\phi) &= - \frac{\lambda_\phi^2}{16\pi^2 E^2} \int_{0}^\infty dp \Bigg[ \left( \frac{m_\phi^2 p}{2\omega} L_2^+(E, p) 
- \frac{4 E p^2}{\omega} \right)\frac{1}{e^{\omega/T}-1} \\ 
&{\hspace{4.0cm}} + \left( \frac{m_\phi^2}{2} L_1^+(E, p) - 4 E p \right) \frac{1}{e^{p/T}+1} \Bigg] ~, 
\end{split}
\end{equation}
where
\begin{equation}
L_1^+ (E, p)  = \ln \left|\frac{4 p E + m_\phi^2}{4 p E - m_\phi^2}\right|, 
\end{equation}
\begin{equation}
L_2^+ (E, p)  = \ln \left| \frac{\left( 2 p E + 2 E \omega + m_\phi^2 \right)\left(2 p E - 2 E \omega + m_\phi^2 \right)}
{\left( -2 p E + 2 E \omega + m_\phi^2 \right) \left( -2 p E - 2 E \omega + m_\phi^2 \right)}\right| \ ,
\end{equation}
and $\omega=\sqrt{p^2 + m_\phi^2}$. Notice that we have added the moduli notation to the arguments of the logarithms to indicate that $V_\phi$ is real (i.e. that only the real part of the logarithms contribute when their argument is negative).
The potential $V_\phi$ changes sign with $T$ for a given scalar mass $m_\phi$. In fact, $V_\phi$  takes the asymptotic form
$V_\phi = - {7\pi^2 \lambda_\phi^2 \epsilon T^5}/({90~ m_\phi^4})$ for $m_\phi \gg T$, and
$V_\phi = {\lambda_\phi^2 T}/({16 ~\epsilon})$ for $m_\phi \ll T$~\cite{Dasgupta:2013zpn,Jeong:2018yts}. 

For a vector mediator, after replacing the mass and coupling constant by $m_V$ and $\lambda_V$, the contribution of the NSI to the thermal potential differs 
from the contribution for a scalar mediator only by a factor of 
2~\cite{Dasgupta:2013zpn,Jeong:2018yts}, 
\begin{equation}
\label{eq:v-Vnp}
V_{\rm NSI}= 2 ~ V_\phi(E, T, m_V, \lambda_V). 
\end{equation}

For sterile neutrino  DW  production with only the standard weak interactions, the maximum of $df_{s}(\epsilon, T)/dT$
in the Std cosmology happens at the temperature~\cite{Dodelson:1993je,Kainulainen:1990ds}
\begin{equation}
 \label{eq:Tmaxstd}
  T_{\rm max}^{\rm Std} \simeq 145 ~ {\rm MeV}~\epsilon^{-1/3}~\Big(\dfrac{m_s}{\rm keV}\Big)^{1/3}~,
\end{equation}
and the value of  $T_{\rm max}$ does not change significantly in the different cosmologies we consider (as shown in App.~(A.1.0) of \cite{Gelmini:2019wfp}).

\begin{figure}[tb]
\begin{center}
\includegraphics[width=.5\textwidth]{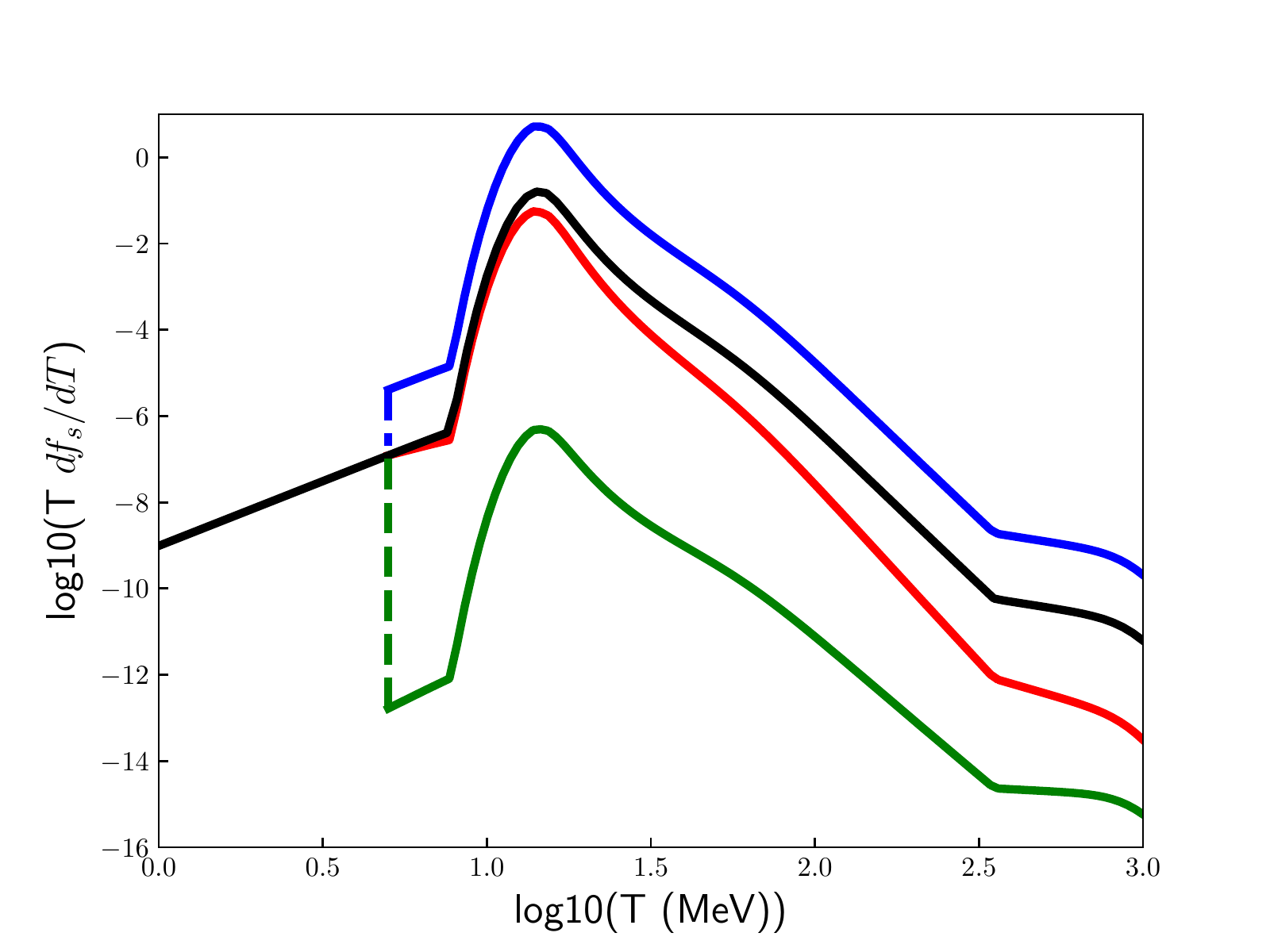}
\caption{Comparison of $T df_{s}/dT$ at $\epsilon=1$ for the Std (black), K (red), ST1 (green), and ST2 (blue) cosmologies for the same scalar mediator model  with $m_\phi = 75$ MeV, and $\lambda_\phi = 10^{-2}$ and a  $m_s = 7.1$ keV sterile neutrino with $\sin^2 2\theta = 5\times10^{-11}$ which could account for the putative 3.5 keV line~\cite{Bulbul:2014sua,Boyarsky:2014jta}. All cosmologies transition to the standard below 5 MeV. A dashed line connects the discontinuous transition of the ST1 and ST2 cosmologies. \label{fig:dfdtgraph}}
\end{center}
\end{figure}

\begin{figure}[tb]
\begin{center}
\includegraphics[trim={0mm 0mm 40 0},clip,width=.475\textwidth]{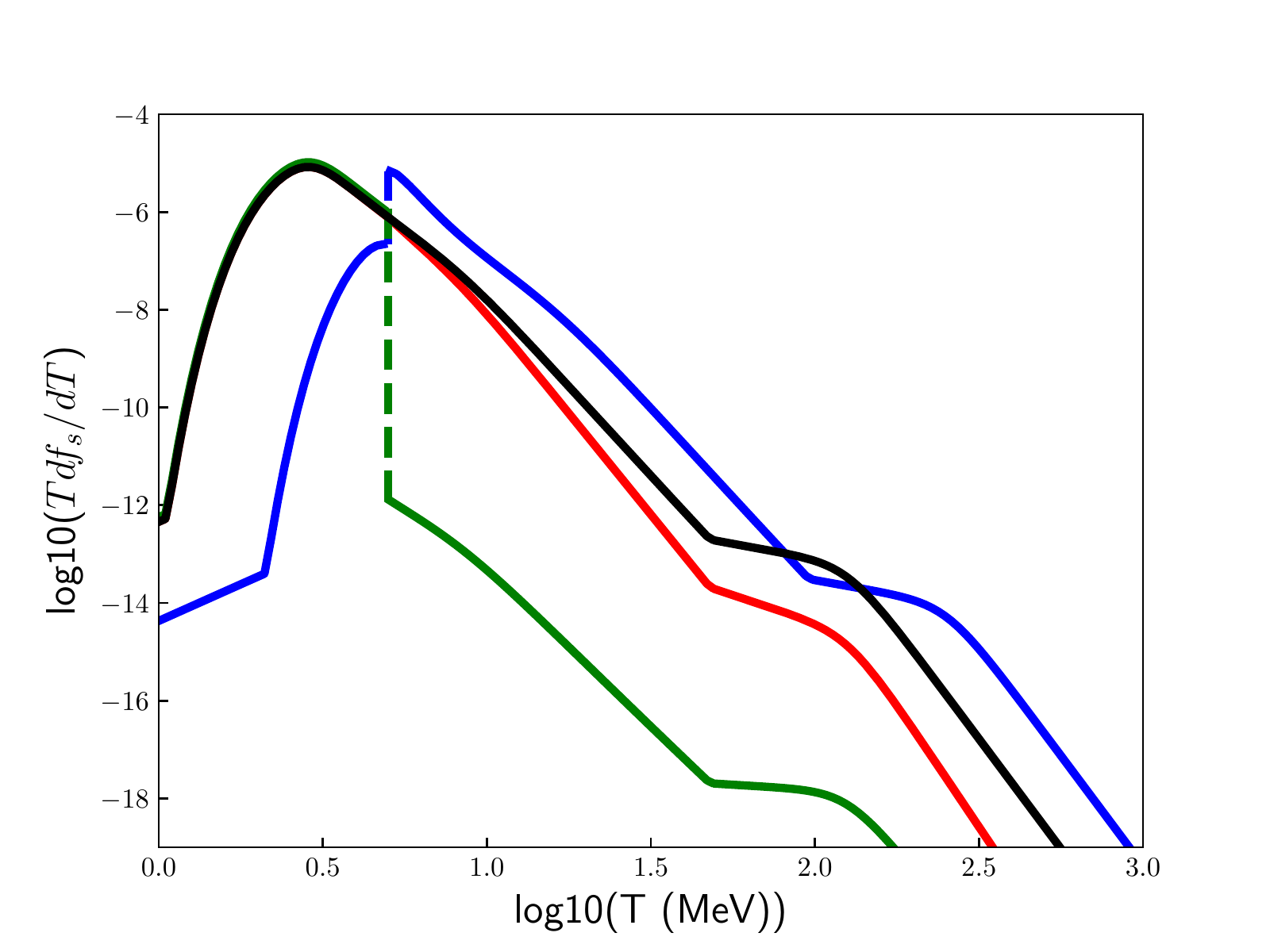}
\includegraphics[trim={0mm 0mm 40 0},clip,width=.475\textwidth]{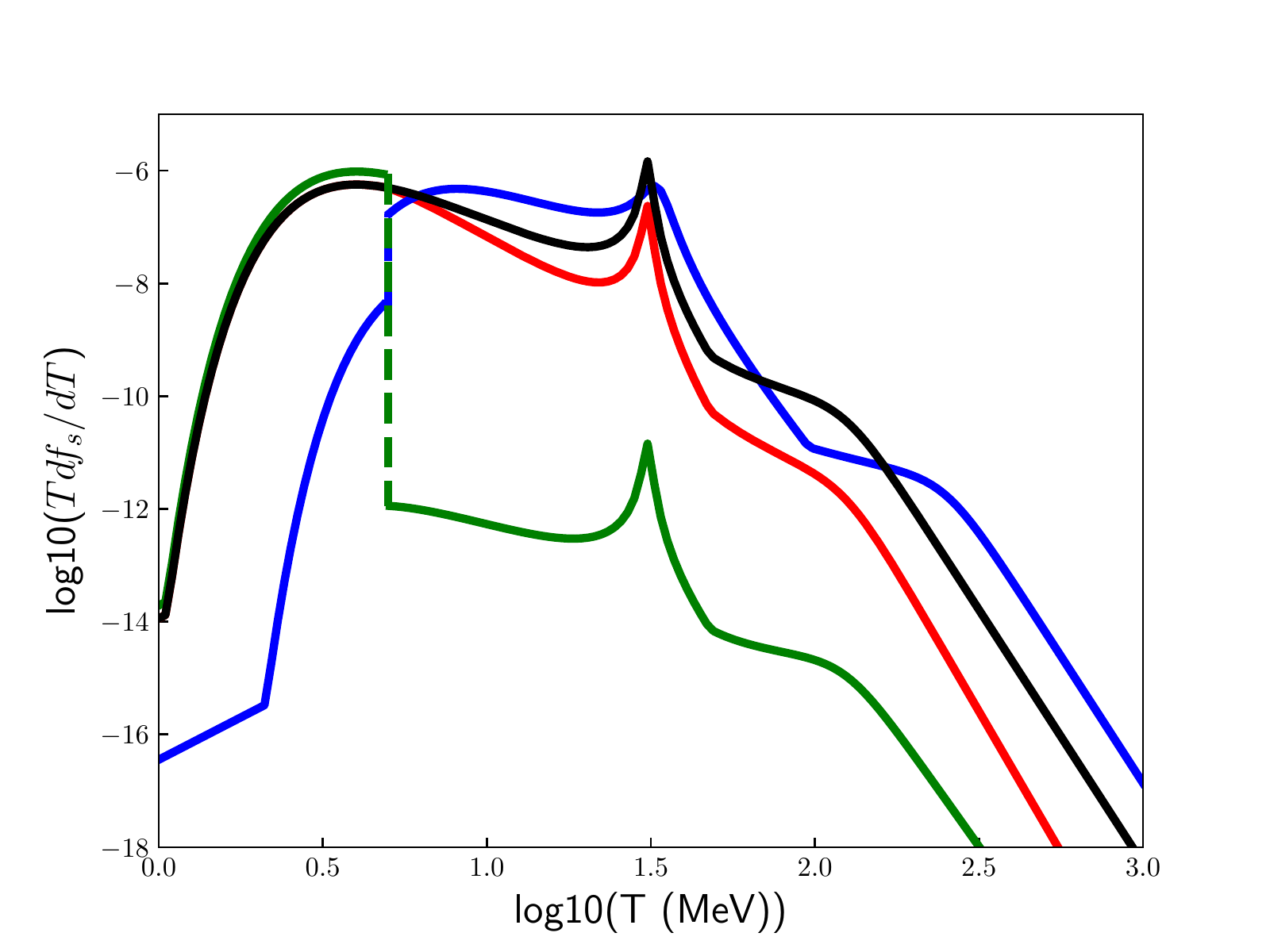}

\caption{  \label{fig:undergraphs} Plots of $T df_{s}/dT$ for $\epsilon=1$ as function of $T$ for two $m_s$ values of the highest producing scalar mediator models, those which determine the mixing angle of the underproduction line (the line below which $\Omega_s < \Omega_{\rm DM}$)  in  the top panel of Fig.~\ref{final-figure}, for the Std (black), K (red), ST1 (green), and ST2 (blue) cosmologies. Dashed lines indicate the sharp transitions we assumed of ST1 and ST2 cosmologies to the standard cosmology (changes would be smoother with more realistic transitions models). \textbf{Left:} 
 For $m_s = 7.1$ keV.  Other parameters are:  $\sin^2 2\theta = 8\times10^{-18}$ for Std and K, $1\times10^{-17}$ for ST1, and $1.2\times10^{-18}$ for ST2; $m_\phi=10\textrm{ MeV}$ for Std, K and ST1, and $20\textrm{ MeV}$ for ST2; and $\lambda_\phi =  1\times 10^{-2}$. \textbf{Right:} For $m_s = 100$ keV. Other parameters are: $\sin^2 2\theta = 2\times10^{-19}$ for Std and K, $3.4\times10^{-19}$ for ST1, and $1\times10^{-20}$ for ST2;  $m_\phi=$10 MeV for Std, K and ST1, and 20 MeV for ST2; and $\lambda_\phi = 1\times 10^{-2}$.  }
 
\end{center}
\end{figure}

The NSI among active neutrinos can both dominate the interaction rate (see Eqs.~\eqref{eq:interaction-off}-\eqref{eq:interaction-B-L-on-shell}) and alter the thermal potential $V_T$ (Eqs.~\eqref{eq:vphi} and~\eqref{eq:v-Vnp}), sometimes creating a resonance. With these additional interactions, the temperature of maximum production rate  can be much lower than in \Eq{eq:Tmaxstd}, as e.g. shown in Fig.~\ref{fig:dfdtgraph} and \ref{fig:undergraphs} in which we plot $T df_{s}/dT$ in the different cosmological models that we consider for a particular scalar mediator and  sterile neutrino model. In these examples the maximum production rate happens at temperatures below $\sim$30 MeV. 

The present sterile neutrino distribution function $f_{s}$ is found by integrating \Eq{eq:boltzmann2} from 
the reheating temperature to the present day temperature.
In practice, we numerically integrate the Boltzmann equation from $T= 10\textrm{ GeV}$ to $T=0.1\textrm{ MeV}$ as most of the production in the models that we consider occurs within this range, 
\begin{equation}
\label{eq:distfunc}
    f_{s}(\epsilon) = \int_{0.1\textrm{ MeV}}^{10\textrm{ GeV}} \frac{df_s}{dT} dT~.
\end{equation}
The number density $n_{\nu_s}$ is obtained by integrating $f_{s}(\epsilon)$ over $\epsilon$, 
\begin{equation}
\label{eq:numden}
      n_{\nu_s}(T_{\nu_s}) = 2\int^\infty_0 \frac{d^3 p}{(2\pi)^3} f_{s}(p) = \frac{T_{\nu_s}^3}{\pi^2} \int^\infty_0 d \epsilon ~ \epsilon^2 f_{s}(\epsilon)~,
\end{equation}
for $T_{\nu_s}< 0.1$ MeV. It is convenient to calculate the present day sterile neutrino number density $n_{\nu_s}$, to express it in terms of the active neutrino number density $n_{\nu_\alpha}$,
\begin{equation}
\begin{split}
\label{eq:numden2}
    n_{\nu_s} =& ~n_{\nu_\alpha}\left(\frac{T_{\nu_s}}{T_{\nu_\alpha}}\right)^3 \frac{\int^\infty_0 d\epsilon \epsilon^2 f_{s}(\epsilon) }{\int^\infty_0 d\epsilon_\alpha {\epsilon_\alpha}^2 f_{\nu_\alpha}(\epsilon_\alpha)} \\
    =& ~ n_{\nu_\alpha} \left(\frac{g_\ast}{10.75}\right)^{-1} \frac{2}{3\zeta(3)}\int^\infty_0 d\epsilon \epsilon^2 f_{s}(\epsilon) ~.
\end{split}
\end{equation}
Here $\zeta(3)\simeq 1.2$ and the present day number density of one active neutrino species is $n_{\nu_\alpha}=112$ cm$^{-3}$. As explained in Sec.~2 we take $g_\ast = 10.75$. Finally, the fraction of the DM consisting of sterile neutrinos is 
\begin{equation}
\label{eq:omegas}
    f_s = \frac{\Omega_{s}}{\Omega_{\rm DM}} = \frac{\rho_{\nu_s}}{\rho_{\rm DM}} = \frac{m_s n_{\nu_s}}{\rho_{\rm DM}}~.
\end{equation}
As usual, here $\Omega= \rho/\rho_c$, $\rho_{\nu_s}$ is the present energy density of sterile neutrinos, the DM density is $\rho_{\rm DM}=0.26\rho_{c}$, $\rho_{c}$ is the  present critical density, $\rho_{c}=1.05\times10^{-5}~h^2~\textrm{GeV/cm}^3$, and we take $h=0.7$.

\section{Constraints and regions of interest}
\label{sec:limits}

To produce  the plots in Fig.~\ref{final-figure}, we discretize and then scan over sterile neutrino mass and mixing  in the ranges $1\textrm{ keV}<m_s<1\textrm{ MeV}$ and $10^{-21}<\sin^2 2\theta <10^{-6}$, and over mediator masses and couplings in ranges suited to obtain $\Omega_{s}= \Omega_{\rm DM}$ and allowed by all limits.

For the scalar mediator model (top panel of Fig.~\ref{final-figure}) we scan over the ranges  $5\textrm{ MeV} < m_\phi < 10 \textrm{ GeV}$ and $10^{-6}<\lambda_\phi<1$,
excluding the  $m_\phi$-$\lambda_\phi$ combinations which violate the bounds from kaon decay~\cite{E949:2016gnh}, neutrino lifetime, and BBN~\cite{Escudero:2019gzq} considerations presented in Fig. 2 of Ref.~\cite{DeGouvea:2019wpf}. 
For the neutrinophilic vector mediator model (middle panel of Fig.~\ref{final-figure}),  we scan over the ranges $5\textrm{ MeV} < m_V < 100 \textrm{ GeV}$ and $10^{-8}<\lambda_V< 1$, excluding the mass-coupling values rejected by bounds from Higgs invisible decays~\cite{PhysRevD.98.030001}, Kaon decays~\cite{PhysRevD.98.030001}, Z boson decays, Supernova 1987A~\cite{Escudero:2019gzq}, and BBN~\cite{Escudero:2019gzq} data shown in Fig.~6 of Ref.~\cite{Kelly:2020pcy}. 
For the $B-L$ vector mediator model (bottom panel of Fig.~\ref{final-figure}), we scan over the ranges $20\textrm{ MeV} < m_V < 100 \textrm{ GeV}$ and $10^{-8}<\lambda_V< 1$,  excluding the $m_V$-$\lambda_V$ values rejected by a combination of experimental limits~\cite{Bauer:2018onh} of several electron and proton beam dump experiments (see Ref.~\cite{Bauer:2018onh} for a full list) and Texono~\cite{TEXONO:2009knm}, Babar~\cite{BaBar:2014zli}, LHCb~\cite{LHCb:2017trq} and CHARM~\cite{CHARM:1985nku} data, shown in Fig. 8 of Ref.~\cite{Kelly:2020pcy}.

To a large extent, the differences in the panels of Fig.~\ref{final-figure} stem from the more restrictive limits on the neutrinophilic and $B-L$ vector mediator mass and coupling compared to the relatively open scalar mediator parameter space. 

For each combination of  discrete sterile neutrino mass and mixing values, we compute $\Omega_{s}$ for all discrete mediator masses and couplings, for each NSI model and cosmology. Since the mediator mass and couplings  are continuous parameters,  if there is a model we compute with $\Omega_{s}>\Omega_{\rm DM}$ and another with $\Omega_{s}<\Omega_{\rm DM}$ for a sterile neutrino mass and mixing, we infer that there is a model with $\Omega_{s} = \Omega_{\rm DM}$.

For the small mixing angles relevant for our work, $\Omega_{s}$ is proportional to $\sin^2 2\theta$, thus for each sterile neutrino mass the underproduction boundary in our plots, the line of the smallest mixing able to produce a DM fractions $f_s=1$ (thus below it there are no models that produce enough sterile neutrinos to constitute the whole of the DM) is determined by the highest producing NSI models. Conversely, the lowest producing NSI models determine the overproduction boundary, the line of the largest mixing angles able to produce $f_s=1$ (the line above which all models produce a density in sterile neutrinos larger than the DM density).

In this manner, we map out the $m_s- \sin^2 2 \theta$ parameter space for all four considered cosmologies, for each mediator model in each of the panels of Fig.~\ref{final-figure}. For each cosmology, sterile neutrinos can account for all of the DM in between the underproduction and overproduction lines, the two solid  black lines for the Std, red lines for the K, green lines for the ST1 and blue lines for the ST2 cosmologies.  For each mediator model we hatch (in green) the regions above the ST1 overproduction line where $\Omega_s > \Omega_{\rm DM}$ in all the cosmologies we consider,  and (in blue) below the ST2 underproduction line,  where $\Omega_s < \Omega_{\rm DM}$ in all the cosmologies we consider. Sterile neutrinos can account for all of the DM in the parameter space in between these hatched region for some combination of cosmology and NSI model. 

We also show in Fig.~\ref{final-figure} the regions excluded by two prominent bounds for keV-mass scale sterile neutrinos, under the assumption of $\Omega_{s} = \Omega_{\rm DM}$: the Tremaine-Gunn bound (orange band) and X-ray constraints (gray regions). 

The Tremaine-Gunn bound is derived from phase space considerations of fermions that compose the whole of the DM~\cite{Tremaine:1979we}. The region shown corresponds to recent improvements to this bound using dwarf spheroidal galaxies which restrict sterile neutrinos with mass $m_s\lesssim 2\textrm{ keV}$ if they constitute all of the DM~\cite{Boyarsky:2008ju}. There are other limits on sterile neutrino DM which extend to larger masses, but are model dependent and thus not included.  Observations of Lyman-$\alpha$ forests restrict the free-streaming scale of warm DM and strongly constrains keV-mass scale sterile neutrinos. Such bounds can limit the sterile neutrino abundance for $m_s<$ 30 keV~\cite{Baur:2017stq}. However, due to their reliance on the free-streaming scale, and thus the DM particle momentum distribution, Lyman-$\alpha$ bounds vary for different NSI models and  cosmologies which can produce much colder (lower average momentum) or hotter (higher average momentum) sterile neutrino distributions than the DW production in the standard cosmology. Since we survey four different cosmologies and a range of NSI models, we do not show the Lyman-$\alpha$ constraints, although the parameter space for keV-mass scale sterile neutrinos is undoubtedly affected to some extent.

\begin{figure}[tb]
\begin{center}
\includegraphics[width=.5\textwidth]{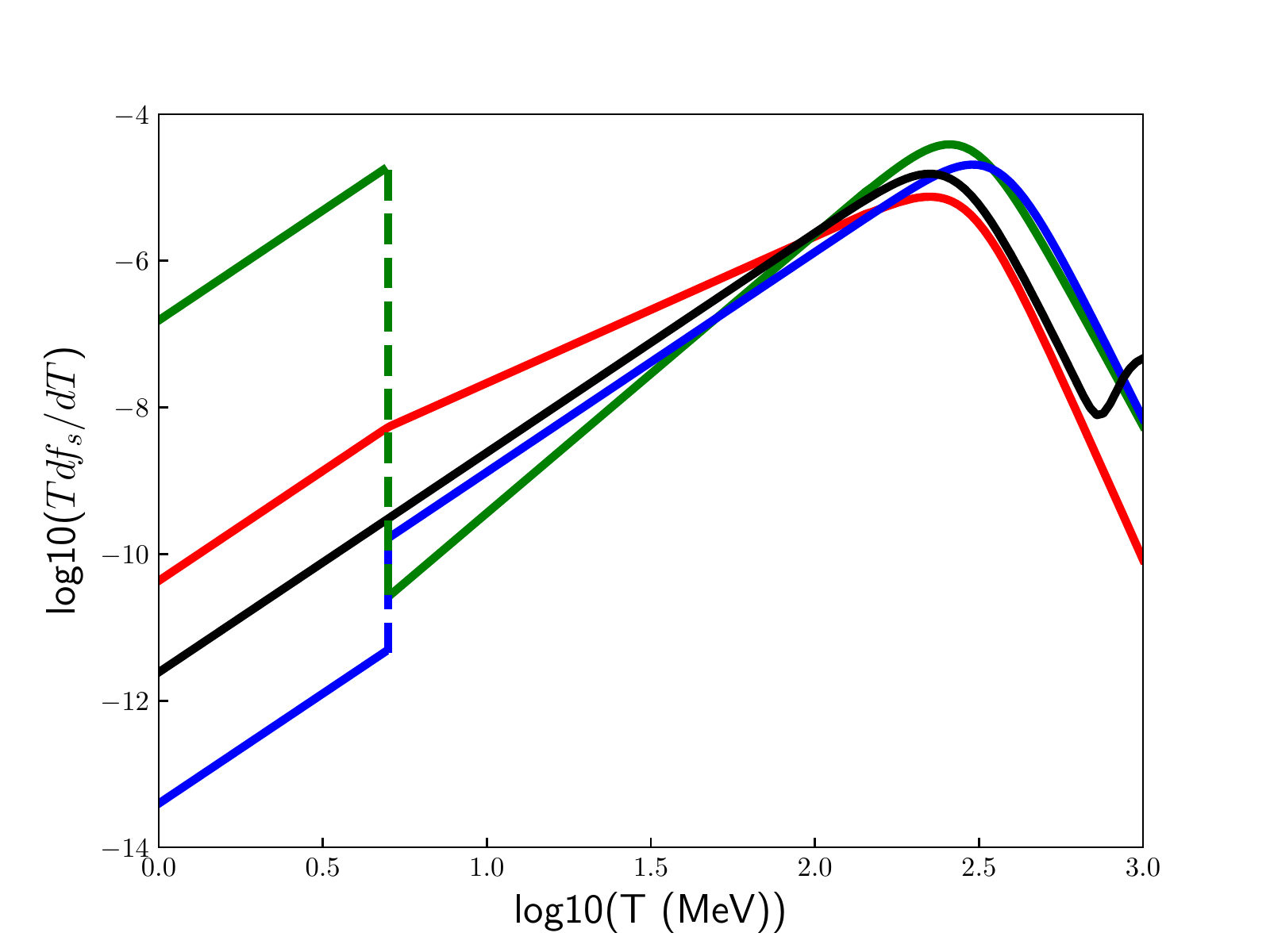}
\caption{Plots of $T df_s/dT$ for $\epsilon=1$ of the lowest producing models, those which determine the mixing angle of the overproduction line (the line above which $\Omega_s > \Omega_{\rm DM}$)  in  the top panel of Fig.~\ref{final-figure}, for $m_s = 7.1$ keV, for the Std (black), K (red), ST1 (green), and ST2 (blue) cosmologies. The other parameters are: $\sin^2 2\theta = 1.6\times10^{-8}$ for Std, $2.8\times10^{-7}$ for K, $1\times10^{-3}$ for ST1, and $2.6\times10^{-10}$ for ST2;  $m_\phi=6.1$ GeV for Std and 12.4 GeV for K, ST1 and ST2; $\lambda_\phi$ is $1\times 10^{-2}$  for Std, $3\times10^{-2}$ for K and ST1, and $1\times 10^{-6}$ for ST2.
\label{fig:overgraph} }
\end{center}
\end{figure}

Sterile neutrinos have a two-body decay mode into an active neutrino and a photon with energy $E_{\gamma}=m_s/2$,
with decay rate $\Gamma_{\nu \gamma} \sim m_s^5 \sin^2 \theta$. X-ray observations of the Andromeda galaxy (M31)~\cite{Ng:2019gch}, the Galactic Center~\cite{Perez:2016tcq}, empty sky fields~\cite{Neronov:2016wdd}, and the diffuse extragalactic background radiation (DEBRA)~\cite{Boyarsky:2005us} strongly constrain the sterile neutrino parameter space for masses above a few keV. A sterile neutrino with $m_s=7.1\textrm{ keV}$ and  assuming  $\Omega_s = \Omega_{\rm DM}$ a mixing $\sin^2 2\theta = 5\times10^{-11}$, could produce a putative 3.5 keV decay line found in the stacked X-ray data of galaxy clusters, as well as M31 and the Perseus cluster~\cite{Bulbul:2014sua,Boyarsky:2014jta}, however this interpretation is under scrutiny (e.g.~\cite{Dessert:2018qih}). 

Note that X-ray bounds from observations of M31, the galactic center and empty sky fields implicitly rely on the existence of galaxies and do not apply to sterile neutrinos which decay before the era of galaxy formation. These small decay times correspond to large masses and active-sterile mixings. DEBRA bounds extend the restricted parameter space to smaller neutrino lifetimes, up to the recombination time. X-ray constraints are inapplicable for sterile neutrinos that decay before recombination. However, a combination of supernova~\cite{Kainulainen:1990bn} and CMB~\cite{Fixsen:1996nj} constraints effectively close the upper right hand corner of the panels in Fig.~\ref{final-figure}. Only a small sliver of the parameter space shown in Fig.~\ref{final-figure} is covered by supernova and CMB constraints, so we do not show these bounds explicitly (see Ref.~\cite{Gelmini:2019clw,Gelmini:2019esj,Gelmini:2019wfp} for more details on these limits).

In  Fig.~\ref{final-figure} we also show the projected reach of two experiments that probe keV-mass sterile neutrinos: the three-year reach from Fig.~\ref{final-figure}~\cite{Mertens:2018vuu}of TRISTAN, the upgrade of the tritium beta-decay experiment KATRIN, in magenta (labeled T), and the projected reach of the ion-trapping experiment HUNTER ~\cite{Martoff:2021vxp} in violet (labeled T).

\begin{figure}[!htp]
\begin{center}
\includegraphics[trim={0mm 9.0mm 0 0},clip,width=.54\textwidth]{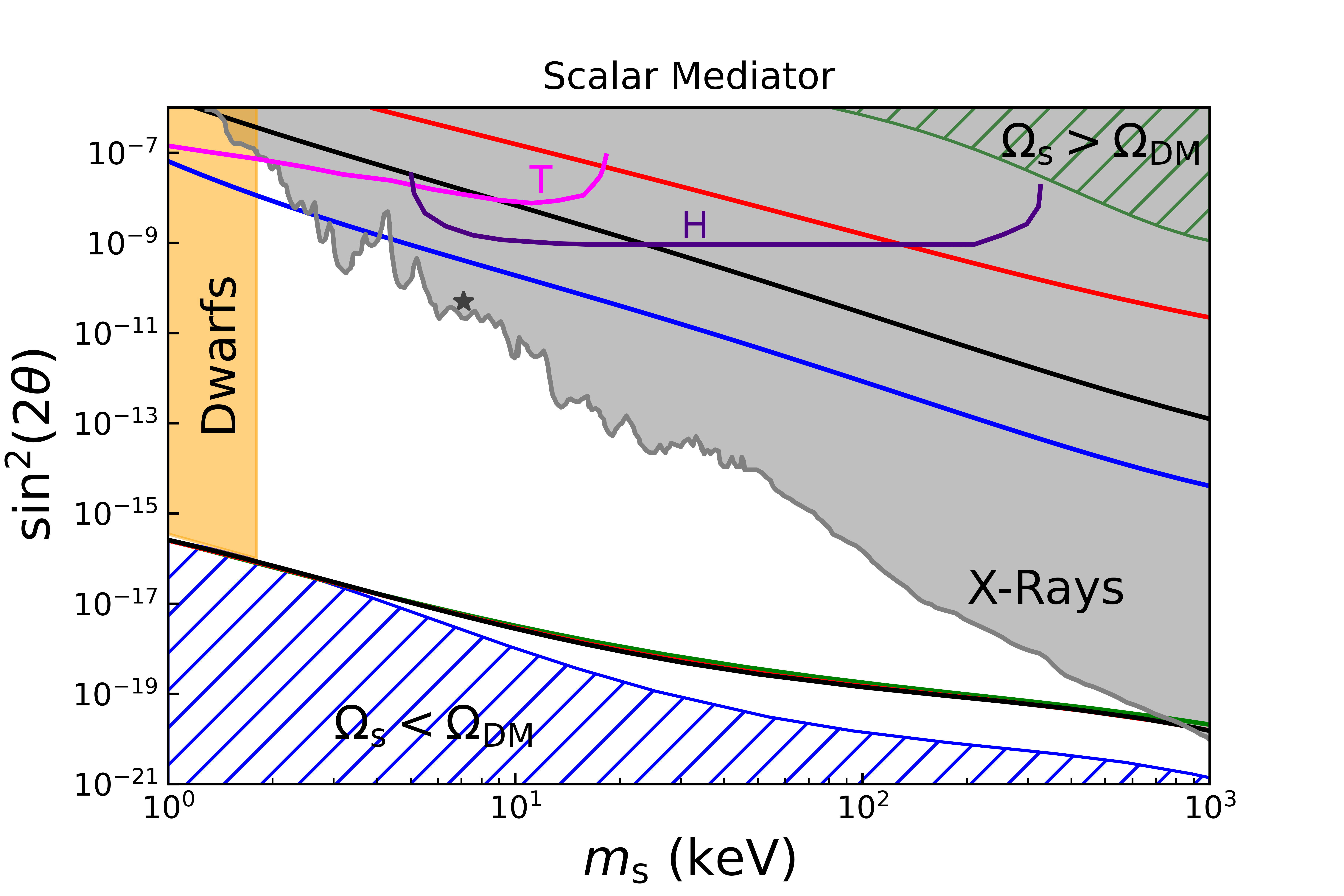}
\includegraphics[trim={0mm 9.0mm 0 3.0mm},clip,width=.54\textwidth]{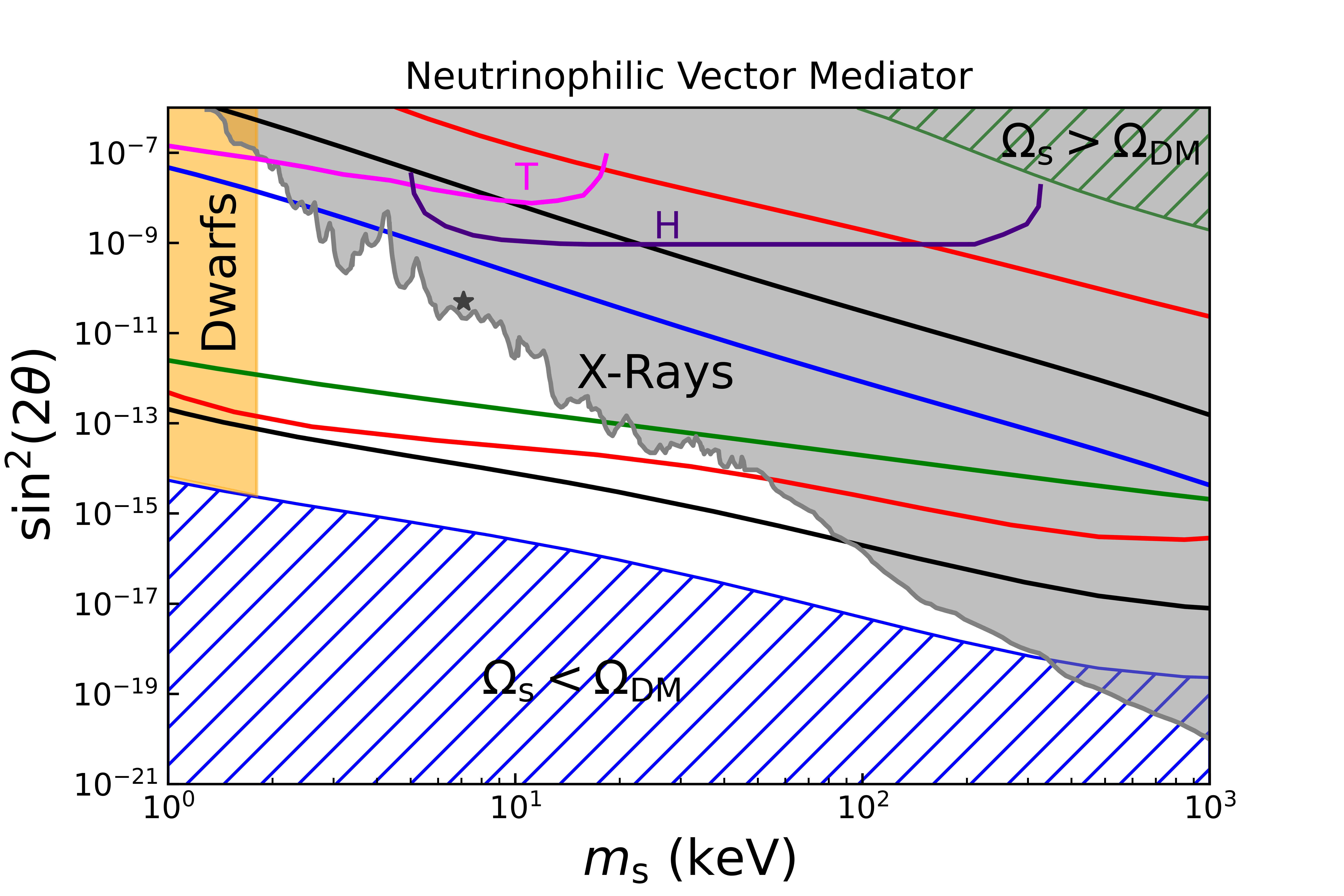}
\includegraphics[trim={0mm 0mm 0 3.0mm},clip,width=.54\textwidth]{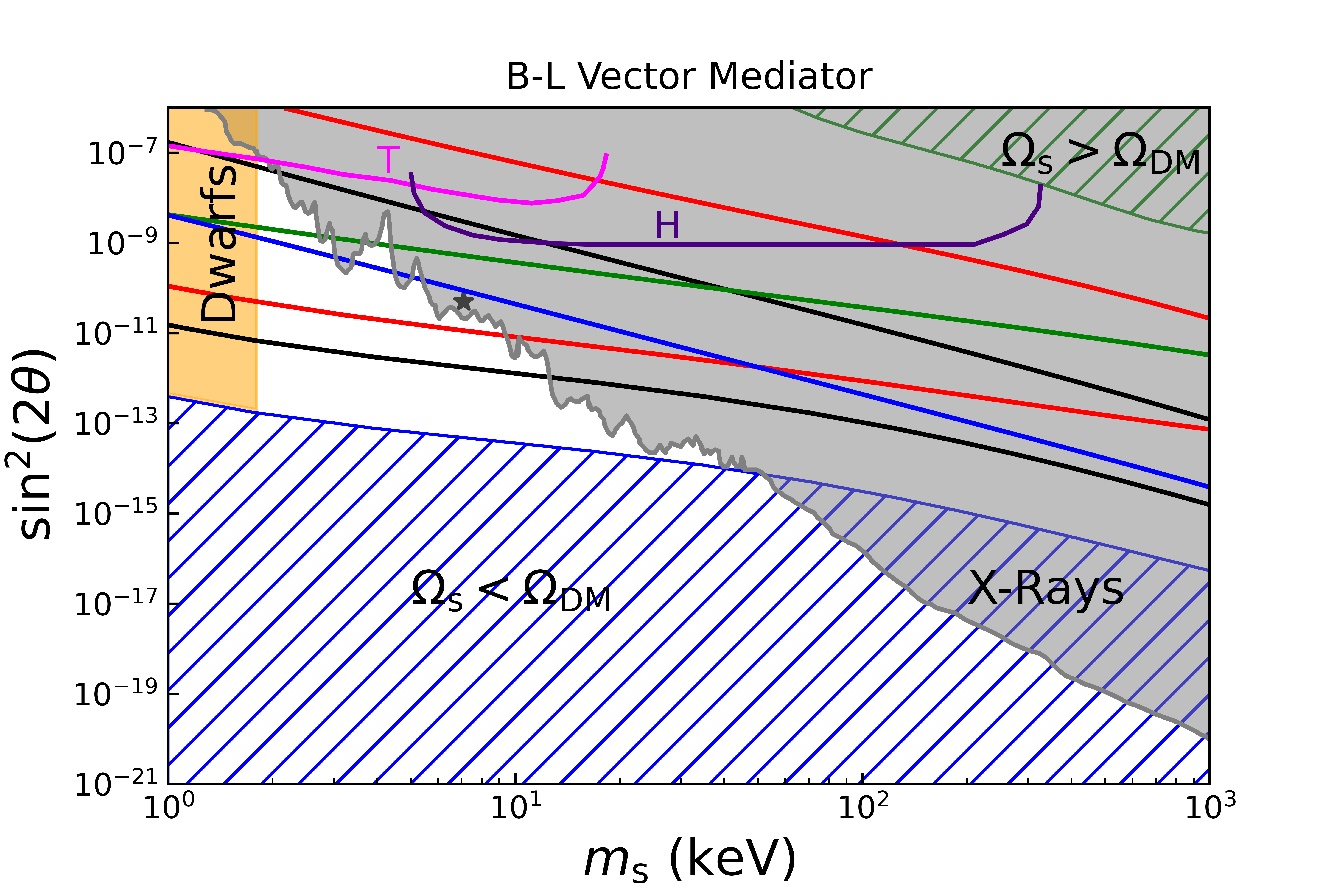}
\caption{\label{final-figure} The regions of mass-mixing where sterile neutrinos can constitute the whole of the DM are in between the two black lines for Std, red lines for K, green lines for ST1, and blue lines for ST2 cosmologies. Also shown are the orange and gray regions ruled out respectively by the Tremaine-Gunn bound applied to dwarf galaxies~\cite{Tremaine:1979we,Boyarsky:2008ju} and X-ray bounds~\cite{Ng:2019gch,Perez:2016tcq,Neronov:2016wdd,Boyarsky:2005us},   the star  at $m_s=7.1$ keV, $\sin^2 2\theta=5\times10^{-11}$ corresponding to the putative 3.5 keV decay signal~\cite{Bulbul:2014sua,Boyarsky:2014jta} (all three assuming $\Omega_s = \Omega_{\rm DM}$), and the projected reach of TRISTAN (T) in magenta and  HUNTER (H) in violet. In the hatched regions, no surveyed combination of cosmology and interactions allow for $\Omega_s = \Omega_{\rm DM}$. 
{\bf Top:} For a scalar mediator.  {\bf Middle:} For a  neutrinophilic vector mediator. {\bf Bottom:} For a $B-L$ vector mediator.} 
\end{center}
\end{figure}

Fig.~\ref{final-figure} clearly shows that our lack of knowledge of cosmology before BBN leads to additional uncertainty with respect to the possible effects of neutrino NSI. For the three NSI mediator models considered, the combination of variations in NSI and cosmology leads to a significant extension of the parameter space allowed by astrophysical limits for sterile neutrinos that comprise all of the DM (for the detailed analysis of the independent impact of uncertainties related to cosmology alone, see Ref.~\cite{Gelmini:2019clw,Gelmini:2019wfp,Gelmini:2019esj,Gelmini:2020duq}). In Fig.~\ref{final-figure} the parameter space of sterile neutrino DM assuming the Std pre-BBN cosmology enclosed between the two black lines is compared with the same regions obtained from making different assumptions.

Cosmologies with higher Hubble expansion rates, such as the K and ST1,  tend to decrease the production rate for the same active-sterile mixing and mass. Since the overproduction boundaries in Fig.~\ref{final-figure} are determined by the models which produce the least amount of sterile neutrinos, K and ST1 models raise the overproduction line to larger mixing angles, an effect that is most pronounced in the ST1 cosmology. Conversely, the ST2 cosmology, which has a decreased expansion rate relative to the Std cosmology, generally increases sterile neutrino production for a fixed mixing and mass, and thus shifts the underproduction lines downwards, thus extending the parameter space where $\Omega_s= \Omega_{\rm DM}$ to smaller mixing angles, where much of the opened region for all three NSI mediator models is free from astrophysical constraints.
 
In the top panel of Fig.~\ref{final-figure}, we see that in the scalar mediator model different cosmologies have a much larger effect on the location of the overproduction lines than   on the location of the underproduction lines. This is because the sterile neutrino production in the models that determine the latter lines happens in large part at $T< $ 5 MeV, where all cosmologies coincide with the standard cosmology, as shown in Fig.~\ref{fig:undergraphs}. Instead the overproduction lines are determined by models in which the maximum of the production happens at larger temperatures, at which the expansion rates are very different in the various cosmologies, as shown in Fig.~\ref{fig:overgraph}.

For keV-mass scale sterile neutrinos, in the Std cosmology DW scenario without any NSI, the production rate peaks at a much higher temperature, $\mathcal{O}(100)\textrm{ MeV}$ (see Eq.~\eqref{eq:Tmaxstd}). With the inclusion of the scalar mediated NSI, the production rate can peak at temperatures close to or lower than the mediator mass,  because the interaction rate in Eq.~\eqref{eq:interaction-on} is maximum for $\omega\simeq1$.  As shown in Fig.~\ref{fig:undergraphs}, for the highest producing models $m_\phi$ is of the order of 10 MeV and the NSI contribute significantly to the production. The overproduction lines are instead  largely determined by the production through weak interactions and not the NSI, thus the temperature of maximum production is not related to the mediator mass.

 From the form of the Boltzmann equation, Eq.~\eqref{eq:boltzmann2}, one can see that the production rate is maximal when a large interaction rate is combined with a small
expansion rate. 
 This explains why models with relatively light scalar masses $\sim 10-20\textrm{ MeV}$ and low peak production temperatures are 
 those with the largest production rates.
Our choice of the transition temperature $T_{\rm tr} =$ 5 MeV, close to the maximum production peak in these models, and our approximation of a transition to the standard cosmology with a sharp jump in the ST1 and ST2 cosmologies (see Fig.~\ref{fig:hgraph}) explains the complicated behavior of the production rates near the transition temperature shown in Figs.~\ref{fig:undergraphs} and~\ref{fig:overgraph}. The abrupt changes of the production rate in the ST1 and ST2 cosmologies would be smoother with a more realistic transition model to the standard cosmology, however the general features we find would not change.

In the top panel of Fig.~\ref{final-figure} the shift  in the ST2 underproduction line  grows as $m_s$ increases.
We see in the left panel of  Fig.~\ref{fig:undergraphs}, that the maximum production rate for small $m_s$ for ST2 is at a temperature of approximately
$10\textrm{ MeV}$, above but close to the transition temperature of $5\textrm{ MeV}$, so that part of the production happens below it, in the standard cosmology. For larger $m_s$ instead, as shown in the right panel of  Fig.~\ref{fig:undergraphs}, more of the production happens above $5\textrm{ MeV}$.
The underproduction lines for the other three cosmologies
in the top panel of Fig.~\ref{final-figure} diverge only very slightly for increasing values of $m_s$. Comparing their production rates for 7 and $100\textrm{ keV}$ (left and right panels of Fig.~\ref{fig:undergraphs}), their peak production temperatures increase weakly with increasing mass, and most of the production still occurs below the transition temperature of 5 MeV.

With respect to the overproduction lines for the scalar mediated NSI,  Fig.~\ref{fig:overgraph} shows that all four models have peak production temperatures in the hundreds of MeV, well above the transition temperature. There is significant production in ST1 below $5 \textrm{ MeV}$ due to the large mixing angles needed to compensate for the much faster expansion rate. However, the bulk of the sterile neutrino production still happens at a few hundred MeV. Thus, there is a direct relation between the change in the Hubble expansion rate relative to the Std cosmology and the shift in the overproduction line.

In contrast to the scalar mediator case, our results for the two vector mediated models are easier to interpret. 

Comparing the exclusion limits on the scalar mediator (Fig.~2 of Ref.~\cite{DeGouvea:2019wpf}) to the corresponding limits for the neutrinophilic and $B-L$ vector mediators (Fig.~6 and~8 of Ref.~\cite{Kelly:2020pcy}, respectively), large $\mathcal{O}(10^{-2})$ coupling constants are allowed for light scalars with mass $m_\phi \gtrsim 5\textrm{ MeV}$ whereas they are forbidden in the vector models. The scalar mediator models with the largest production have $\lambda_\phi \simeq 10^{-2}$. Due to the limited parameter space of the vector mediator models, the maximally producing models which determine the underproduction line are not those with light vector mediators because only small couplings are allowed for them. Rather, they are models with mediator masses close to or larger than 100 MeV  for which larger couplings are allowed. Consequently, for vector mediated NSI, the  temperature of maximum production is much above the transition temperature of $5\textrm{ MeV}$.

For neutrinophilic vector mediated NSI the  temperature of maximum production ranges
from about $10\textrm{ MeV}$ for $m_s \simeq 1\textrm{ keV}$ to a few hundred MeV for $m_s \simeq 1\textrm{ MeV}$. The full effect of the ST1 and ST2 cosmologies on the underproduction line is evident in the middle panel of Fig.~\ref{final-figure}. The slow departure of the K cosmology underproduction line as $m_s$ increases from 1 keV is due to its gradual transition to the standard cosmology. The K expansion rate at $T=10\textrm{ MeV}$ is modified with respect to the standard only modestly by a factor of 2. This behavior of the K underproduction line is not present for the $B-L$ vector mediator because the available parameter space in the mediator  mass-coupling  plane is even more restricted than for the neutrinophilic vector mediator (see Fig. 8 of Ref.~\cite{Kelly:2020pcy}) and disallows the lightest mediator masses $m_\phi<20$ MeV. As a result, even for $m_s=1\textrm{ keV}$, the highest producing models have the peak sterile neutrino production at temperatures around $50\textrm{ MeV}$. The very limited $m_\phi$-$\lambda_V$ parameter space for the $B-L$ vector mediator severely limits the effects of this NSI on sterile neutrino production, so the corresponding $m_s$-$\sin^2 2\theta$ parameter space of sterile neutrino that can account for all of the DM, as shown in the lower panel of Fig.~\ref{final-figure},  is considerably smaller than in the scalar and neutrinophilic vector mediator cases.

Finally, let us comment on sterile neutrinos which constitute only a part of the DM. Notice that since $f_s$ is proportional to $\sin^2 2\theta$ for the small mixing in our figures, using the plots in Fig.~\ref{final-figure} it is easy to find the regions where sterile neutrinos would constitute any fixed fraction $f_s <1$ of the DM, i.e. where $\Omega_{s} = f_s \Omega_{\rm DM}$ with $f_s$ fixed to any value smaller than 1: the region shown for $\Omega_{s} = \Omega_{\rm DM}$ in Fig.~\ref{final-figure} would move down in our plots by a factor $f_s$ in $\sin^2 2\theta$.

Notice that for fixed $f_s <1$,  the region forbidden by X-ray observation would move upwards.  The boundary of the region forbidden by X-rays observations corresponds to a fixed X-ray flux, i.e. it is proportional to the sterile neutrino density times decay 
rate, and this product is independent of $\sin^2\theta$. Thus, if the sterile neutrino density decreases by $f_s$, the X-ray upper limits move upwards towards larger mixings by a factor of $1/f_s$ in $\sin^22\theta$.

\section{Conclusions}
\label{sec:summary} 

Sterile neutrinos constitute an attractive DM candidate and have direct connections to other puzzles, such as the nature of neutrino masses. While stringent astrophysical bounds have been placed on keV-mass sterile neutrino DM, i.e. sterile neutrinos that constitute all of the DM, the viable parameter space is subject to significant theoretical uncertainties. One such source is that of possible additional neutrino interactions (NSI) that could appear in minimal models. Another source is the lack of knowledge about the  pre-BBN cosmological history, which in some well motivated scenarios could be very different  than typically assumed. These effects can separately dramatically increase the viable parameter space in which sterile neutrinos can account for all of the DM abundance.

To identify the maximal available parameter space for sterile neutrino DM, we have analyzed the impact of the combined uncertainties associated with additional neutrino interactions as well as early Universe pre-BBN history on the cosmological sterile neutrino production and the resulting constraints. We consider several representative scenarios of active neutrino NSI, including scalar and vector (neutrinophilic and $B-L$ models) mediated NSI, as well as scalar-tensor (ST1 and ST2 scenarios) and kination (K) cosmologies that allow for a significant variation in the Universe expansion rate.

Compared to previous results in the literature for scalar mediated NSI assuming the standard pre-BBN cosmology, we identify a large extension of sterile neutrino DM parameter space for higher mixing angles in the context of K and ST1 cosmologies. Due to
effects of the transition between the early pre-BBN cosmology and the known cosmology of the late Universe, for the scalar mediated NSI the extension of the sterile neutrino DM region to smaller mixing angles is not as sizeable but is still significant. The reason is that the models responsible for the lower boundary of the sterile neutrino DM region produce most of the sterile neutrinos at temperatures below the transition to the standard cosmology, for which all cosmological models coincide.  While the newly available parameter space at high mixing angles is forbidden by X-ray bounds, we have identified an extended viable region below X-rays limits in the context of ST2 cosmology.

Compared to previous results in the literature for neutrinophilic vector mediated neutrino NSI, we find a significant extension of the viable parameter space for sterile neutrino DM to both higher mixing angles (for the ST1 and K cosmologies) and also to smaller mixing angles (for the ST2 cosmology). Unlike in the scalar mediator case, the temperature of maximum production rate in these models is always well above the transition temperature of $5\textrm{ MeV}$ associated with known late Universe cosmology, so that the underproduction lines for the different cosmologies do not coincide with that of the standard cosmology. 

Cosmological effects on sterile neutrino production are even more pronounced in the case of $B-L$ vector mediated NSI. Due to the more restrictive laboratory bounds on the viable vector boson mediator mass and coupling for this model, the effects of NSI on the available sterile neutrino DM parameter space are limited. In the standard cosmology, only a small portion of space is not forbidden by either X-ray or dwarfs astrophysical bounds. Relative to the narrow viable region for sterile neutrino DM if the standard cosmology is extrapolated to the pre-BBN Universe, the region is considerably extended for the ST2 cosmology below the upper bounds from X-ray observations.

Although the projected reaches of the TRISTAN and HUNTER laboratory experiments are forbidden by X-ray bounds for sterile neutrino DM, they would evade these limits for sterile neutrinos that are sufficiently underdense (when sterile neutrinos constitute a fraction $f_s$ of the DM, the X-ray bounds are shifted upwards by a factor $1/f_s$). For the particular NSI cases considered here, still one would have to appeal to the uncertainties associated with the unknown pre-BBN cosmology in order to have sterile neutrinos within the reach of these experiments and  within the regions of constant density as we discuss, which would move downwards by a factor $f_s$ (the reach of TRISTAN and HUNTER can be freed of any astrophysical bounds by considering the effects of uncertainties associated with early pre-BBN cosmologies alone, without NSI).

As we have demonstrated,  theoretical cosmological uncertainties can significantly impact the viable parameter space for sterile neutrino DM, besides theoretical uncertainties due to non-standard neutrino interactions, and thus must be taken into account when interpreting experimental and observational results.

\acknowledgments
\addcontentsline{toc}{section}{Acknowledgments}
 
We thank Manibrata Sen for useful comments on interaction rates. The work of G.B.G., P.L. and V.T. was supported in part by the U.S. Department of Energy (DOE) Grant No. DE-SC0009937. 
P.L. was also supported by the Grant Korea NRF-2019R1C1C1010050. V.T. was also supported by the World Premier International Research Center Initiative (WPI), MEXT, Japan. This work was performed in part at the Aspen Center for Physics, which is supported by the National Science Foundation grant PHY-1607611.  

\bibliography{sternumodcos}
\addcontentsline{toc}{section}{Bibliography}
\bibliographystyle{JHEP}
\end{document}